# A Methodology and Software Architecture to Support Explainability-by-Design

Trung Dong Huynh, Niko Tsakalakis, Ayah Helal, Sophie Stalla-Bourdillon, and Luc Moreau

**Abstract**—Algorithms play a crucial role in many technological systems that control or affect various aspects of our lives, from credit assessment to exam grading, medical screening, criminal justice, and more. As a result, providing explanations for their decisions to address the needs of users and organisations is increasingly expected by laws, regulations, codes of conduct, and the public. However, as laws and regulations do not prescribe how to meet such expectations, organisations are often left to devise their own approaches to explainability, inevitably increasing the cost of compliance and good governance. Hence, we envision *Explainability-by-Design*, a holistic methodology characterised by proactive measures to include explanation capability in the design of decision-making systems. The methodology consists of three phases: (A) Explanation Requirement Analysis, (B) Explanation Technical Design, and (C) Explanation Validation. This paper describes its software engineering phase (B), a technical workflow to implement explanation capability from requirements elicited by domain experts for a specific application context. Outputs of this phase are a set of configurations, allowing a reusable explanation service to exploit logs provided by the target application to create provenance traces of the application's decisions. The provenance then can be queried to extract relevant data points, which in turn can be used in explanation plans to construct explanations personalised to their consumers. Following the workflow, organisations will be able to design their decision-making systems to produce explanations that meet the specified requirements, be it from laws, regulations, or business needs. To facilitate the process, we present a software architecture with reusable components to incorporate the resulting explanation capability into an application. Finally, we applied the workflow to two application scenarios in a case study and measured the associated development costs. It was shown that the proposed approach to construct explanations is tractable in terms of development time, which can be as low as two hours per sentence.

✦

## 1 INTRODUCTION

THE 1990s saw a movement of privacy-enhancing technologies [1] aimed at addressing increasing requirements for data privacy and, by extension, data protection. It led to the concept of "Privacy by Design" [2] and "Privacy by Default", which has been now adopted by key legal frameworks [3]. In the same vein, the requirements for explainability to address the needs of users and organisations are now being recognised widely [4], [5]. Increasingly, laws, regulations, and codes of conduct put a duty on organisations that make automated decisions about their customers or users to provide them with an explanation of the 'logic' associated with such decisions. It has been dubbed the 'right to explanation' [6]. However, relatively few technical recipes are available to help build systems that offer the level of explainability required [7]. Hence, by analogy to "Privacy by Design", we envision *Explainability-by-Design* (EbD), a methodological approach that tackles this issue in the design and architecture of IT systems and business practices. EbD is characterised by proactive measures to include explanations in the design rather than relying on ad-hoc measures to bolt on explanation capability as an afterthought or add-on. It advocates for explainability to be integral to the decision-making system, enriching its functionality with capabilities that can address regulatory requirements but also functional and business needs. As a result, the system's behaviour is said to be explainable by design.

Here, we take a technical view in which a *decision* can be any output of a data processing pipeline produced after considering one or more data points, rather than a legal view of a decision (which often includes considerations of the effect it has on a person [8]). We define an *explanation* to be a statement providing details or reasons that relate to or underlie a decision to enable the recipient of the statement: (i) to better understand the decision or (a particular aspect of) the decision-making process (e.g. its process, data basis, influences, responsibility); and where necessary (ii) to take action (e.g. contest a decision, correct the decision-making process). Explanations, therefore, must be sufficiently meaningful to provide relevant stakeholders with an understanding of how a decision was made and, specifically to end users, to enable them to exercise their legal rights [9].

Against the impetus to explainability and the rising popularity of decision-making applications built on machine learning (ML), there has been a resurgence of research in Explainable Artificial Intelligence (XAI) [10]. A key focus of XAI is on revealing the relationships between an ML model's input data and its prediction or classification, either at a global level (i.e. understanding its whole logic, such as permutation feature importance [11]) or at a local level (i.e. why the model made a specific prediction for an instance, such as SHAP [12], LIME [13]). ML models, however, are typically only one part of much wider pipelines and processes that influence decision-making. While learning which data point caused a particular recommendation by an ML model can be insightful to a data engineer, it is unlikely to help a layperson understand how a decision was reached (and what can be done about it). According to Miller, explanations are not just the presentation of associations and causes, but they are contextual [14]. To be sufficiently meaningful to users and organisations, explanations cannot just focus on the 'black box' but also have to provide information about events that have influenced a decision's outcome before and after the application of the algorithmic models [15], [16]. On this front, Chazette and Schneider [5] identified various aspects of explainability requirements that should be considered during the requirements engineering process of a software application. Once such requirements are elicited and analysed, the taxonomy of explanations proposed in [17] can serve as a guide to determine the contents of explanations to produce according to the application's specific needs.

Having specified the required explanations and designed their contents, the application's owner is left with the challenging task of producing the explanations for its decisions. The job, at a minimum, includes extracting the relevant data from the application's business logic, constructing explanations from them, and integrating the explanation generation capability into the application. To that end, we tackle the following questions in this work:

1) What are the concrete steps to transform an application's requirements for explanations and their designs into its integrated software functionality?
2) Which software components underpinning the above explanation functionality can be reused to alleviate the cost of implementing explanations, and how can it interface with the business logic?

In addressing the above, this paper proposes an engineering workflow for implementing explanations from their requirements and integrating the functionality into an application in a reliable, repeatable, and reusable manner. It makes use of *provenance*, a form of knowledge graph providing an account of what a system performed to produce, say, a decision, as the means to expose the application's business logic to support explanation generation (see Sec. 5.1 for more details). The workflow specifies how the provenance should incorporate the data needed to produce explanations and how they can be retrieved and transformed into explanations in a natural language, ready to be presented to their target audience. It is part of the Explainability-by-Design methodology, which consists of three phases: (A) Explanation Requirement Analysis, (B) Explanation Technical Design, and (C) Explanation Validation. The workflow is to be carried out by a data engineer in Phase B to implement the explanation requirements previously elicited in Phase A by domain experts within an application's socio-technical context. Application stakeholders then evaluate the produced explanations in Phase C. This paper focuses on the engineering steps and the software architecture to translate the provided requirements into software capability to generate the explanations they specify. Hence, eliciting explanation requirements and evaluating the resulting explanations are not in scope; see [17], [18] for discussions on these topics.

In particular, our contributions are as follows:

- A detailed description of the Explanation Technical Design phase (B), a software engineering workflow that takes the requirements for explainability as input and produces a stand-alone explanation service to receive application data and generate explanations. We describe the concrete steps in each task and specify the inputs and outputs required.
- A software architecture for an explanation service: a reusable architecture with clearly defined interfaces taking the artefacts produced by the above engineering tasks to provision explanation capabilities for a decision-making application. A reference implementation is provided for two different application scenarios.[1]
- A case study of the enactment of the engineering workflow over two application scenarios to evaluate and characterise the technical efforts required to support explanations. The study shows that the approach is tractable in terms of development time and that some technical artefacts can be reused to reduce the overall engineering effort.

In the remainder of the article, we first discuss the related work (Sec. 2) and present the two application scenarios in which the EbD methodology was enacted in the case study (Sec. 3). We overview the three phases of the EbD methodology (Sec. 4) and describe the tasks comprising its Explanation Technical Design phase (Sec. 5). Sec. 6 presents the software architecture for provisioning explanation capabilities, and Sec. 7 discusses the technology chosen for its reference implementation. Sec. 8 assesses the efforts incurred when executing the Explanation Technical Design phase on the two application scenarios. Finally, in Sec. 9, we conclude the article with extension points for future work.

## 2 RELATED WORK

The need for explanations and justification of computer-generated advice is not new [19], but with the recent prevalent adoption of automated and machine-assisted decision-making systems, the public's attention on explanations of their decisions has become widespread [20]. Recognising this, in the EEA and the UK, the GDPR[2] includes a right to access to "meaningful information about the logic involved, as well as the significance and the envisaged consequences of such processing" of personal data [8, Article 15]. A key challenge with automated decision-making systems, however, lies in the technical implementation of such obligations given their typical complexity. The academic community has also recognised the importance of tackling the above explainability concerns, which have become an active research topic worldwide. Several international events are dedicated to this topic: ACM Conference on Fairness, Accountability, and Transparency (ACM FAccT), the Explainable AI workshop at International Joint Conferences on Artificial Intelligence (IJCAI); the Fairness, Accountability, and Transparency in Machine Learning (FAT/ML) workshop[3]. To combat the opaqueness of ML models, a variety of techniques have been proposed (see [21] for an overview, and [10], [22] for extensive surveys). Approaches include designing the learning process to ensure the interpretability of results, approximating a learned model in a more readily intelligible substitution, and offering tools to interact with the model to get a sense of its operation.

---

1. Available online at https://explain.openprovenance.org.
2. The General Data Protection Regulation [8].
3. ACM FAccT conference: https://facctconference.org. FAT/ML workshop: https://www.fatml.org.

In software engineering (SE), ML techniques are also being applied to improve software development; and in some cases, explanations are offered to help the developers understand why a given recommendation is made [23]. For instance, neural networks are employed to identify self-admitted technical debts from code comments and terms picked by the model are shown to help explain the intuition behind the model's decisions [24]. In another instance, explanations are offered for refactoring solutions to help users understand how the solutions are generated [25]. With the plethora of explainability approaches developed for various types of ML algorithms, some techniques were found more suitable for certain explainability goals than others by users [26]. Even so, the main focus of the aforementioned work is on explaining the inner workings of the algorithms employed in decision-making systems and often fails to account for influential events outside the black box [27]. Organisations and end users, for instance, might not be fully aware of the exact practices that generate the training data upon which ML models are built [28].

In contrast, Explainability-by-Design builds on several concepts analogue to the seven foundational principles of Privacy by Design [2] in a holistic approach.

- **Proactive**: EbD does not wait for a request for explanations to investigate forensically how a decision was reached.
- **Explainability by default**: EbD seeks to deliver the maximum degree of explainability, ensuring that all decisions are potentially explainable without requiring any action from the user.
- **Explainability embedded in the design**: EbD is not bolted on as an add-on but is embedded in the design of systems so that it becomes an essential component delivering functionality for the application.
- **End-to-end Explainability**: EbD is applied to all information flows during the entire lifecycle of the data involved.
- **Visibility and Transparency**: EbD seeks to assure all stakeholders are provided with explanations addressing their needs. Explanations themselves can be audited and potentially independently verified.
- **User-centric**: EbD seeks to help end users understand a decision and take action where applicable.

Hence, EbD advocates for the systematic recording of the provenance of decision-making within an application [29], i.e. the audit trails including references to people, data sets, and organisations involved in a decision, along with attribution of data and data derivations. By so doing, an application exposes the decision-making processes that surround a "black-box" model, the actors and the data that influence its decisions, making its business logic transparent by default. Such audit trails provide valuable information from which explanations of how the system arrived at the decision can be constructed to address the what, why, and how questions mostly asked by software users [5]. Those audit trails support *ex-post* explanations, which are constructed after the corresponding decision is made and can refer to the specific data and processing of that decision (as opposed to *ex-ante* generic explanations for informed consent) [30]. Similarly, the provenance of loan decisions was previously shown to be able to support various GDPR-related explanations for such decisions [31]. Such explanations were later demonstrated to be effective means as both internal detective controls and external detective controls, helping organisations meet their accountability and data protection-by-design obligations [18], [32].

This work follows a sequence of methodological approaches for building systems with provenance capabilities. PRIME, a provenance incorporation methodology [33], was defined as a stand-alone methodology to elicit provenance requirements and design applications that surface the appropriate flows so that they can be exported in a provenance description language (which preceded the PROV Data Model [34]). EbD could be seen as an instantiation of PRIME, which aims to surface internal data elements, to expose provenance, which in turn is used to generate explanations. The Provenance Template [35] approach allows provenance to be specified declaratively, using a template with placeholders composed at design time, to be instantiated with runtime values. In the reference implementation of the EbD methodology (Sec. 7), provenance templates were adopted to assist with the Provenance Modelling task. UML2PROV [36] conceptually combines provenance templates with the Unified Modelling Language (UML): it is an adjunct methodology to UML, able to leverage suitably annotated UML diagrams in order to record runtime values required to instantiate provenance templates constructed from such diagrams. All three methodological approaches, PRIME, Provenance Templates, and UML2PROV, are intended to address a broad set of requirements for provenance [37]. Here, EbD can be seen as a methodology focusing on a specific subset of requirements, namely those requiring applications to be explainable. It specifically addresses the explainability and transparency requirements of software systems [5], one of the key software quality characteristics desired in AI-based systems [38], [39].

With generative AI making significant progress in text generation, it is worth examining why an alternative approach was adopted here. A key driver for our explanation generation is the legal obligation that organisations have to comply with in terms of explaining the decisions they make. Such explanations need to be precise, have a specific purpose, and cannot mislead their recipients. Generative AI, however, suffers from issues such as producing statements that are not true or suffering from bias [40]. Therefore, we rely on a traditional approach to natural language generation (NLG) [41] allowing for the sharing and reuse of architectural components, such as a realisation engine [42].

Finally, the EbD methodology is broadly aligned with the human-centred focus for designing explanations [5], [16], [38], [43] by first defining the socio-technical context in which an application is situated, eliciting explanation requirements from the application's stakeholders within the said context in its Phase A (see Sec 4 for more details). The explanations generated are later validated with the stakeholders against the elicited requirements in Phase C. Similar to the step-by-step management's guide to explaining decisions made with AI by the UK's data regulator [7], the methodology's SE phase presents its tasks in a step-wise workflow to help engineers convert requirements for explanations into software capability.

## 3   EXPLANATION SCENARIOS

To illustrate the EbD methodology's steps, we describe two concrete application scenarios in this section. The two scenarios were co-created with our collaborators, a credit rating agency and a local education authority respectively. They will later serve as case studies [44, Ch. 5] for the assessment of the methodology (Sec. 8).

### 3.1   Credit Card Assessment

Credit card applications nowadays are typically assessed by automated systems, taking into account the applicant's credit history provided by a credit referencing agency (CRA). Such applications are often approved or rejected within seconds, without human intervention. In this scenario, an applicant applies for a new credit card with a bank. The bank employs a CRA to search the applicant's credit history and provide a credit report. Such reports typically contain a credit score, representing the applicant's overall creditworthiness, in addition to a variety of credit-related records: past payments (to existing credit agreements), addresses, electoral roll, etc. Assuming that the bank rejects an application, the applicant is entitled to explanations for the bank's decisions according to the prevailing regulations.

### 3.2   School Allocation

Over a million parents apply for school places in primary and secondary schools for their children in the UK every year. Local education authorities routinely employ semi-automated school application systems to help manage the significant number of applications by matching parents' school preferences against a wide variety of ranking criteria set by individual schools, allocating available school places according to such criteria. School allocation decisions are communicated to the parents, informing which school their children are to be admitted to, with little information on the individual circumstances of a child that influenced the decision. Parents who do not get their first choices naturally want to understand the reason, typically resulting in a considerable volume of calls and inquiries from disappointed parents following the national offer day. Hence, producing decision letters tailored to the individual circumstances of a child's application will help parents understand better why their children are allocated to a particular school (and why not to their first choice). At the same time, such targeted explanations should reduce enquiries on their allocation decisions to local authorities and would also assist operators at call centres with providing targeted answers.

## 4   EXPLAINABILITY-BY-DESIGN METHODOLOGY

The EbD methodology is a socio-technical process involving application stakeholders, domain experts, and data engineers. It provides organisations with a recipe to build explanation capability into their decision-making systems while addressing regulatory and business requirements, such as those outlined in the above scenarios. It identifies the key stakeholders involved in the process and defines clearly the inputs and outputs of each task[4] (whose definitions are

---
4. The inputs and outputs of tasks in the EbD methodology, or Artefacts, are highlighted in the texts with the purple colour.

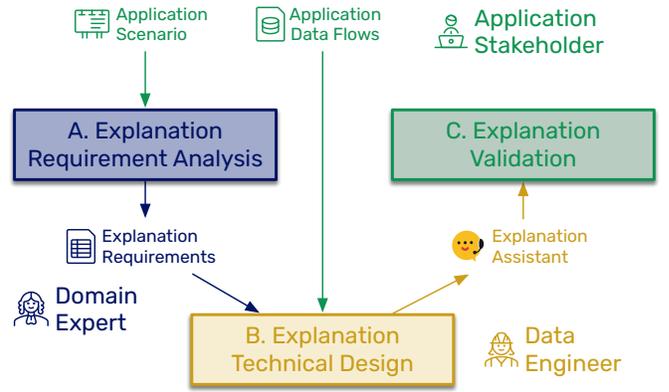

Fig. 1. Overview of the Explainability-by-Design methodology.

provided in Table 2). The EbD methodology, shown in Fig. 1, comprises three phases:

A. **Explanation Requirement Analysis**: From an Application Scenario provided by its stakeholders, the requirements of explanations for the application are captured by the domain expert analysing the prevailing regulations and business needs applicable to the scenario. Depending on the mandates for explanations, the domain expert's role could include business management, lawyers, policymakers, or sociologists. This phase aims to enumerate and characterise all the explanations as mandated by regulations or business needs applicable to decisions made by the application. Collectively, the output from this phase is called the *socio-technical specification of explanations* for the chosen application, referred to as the Explanations Requirements for brevity. It consists of a list of explanations to be supported along with their (technical) characteristics (described below).

B. **Explanation Technical Design** (ETD): This phase consists of a set of technological steps, ingesting the Explanations Requirements from the previous phase and the application's Data Flows to construct an explanation capability that a system's triggers can activate to generate the required explanations. At a high level, the ETD phase produces an Explanation Assistant service that generates explanations according to the Explanations Requirements. It consists of four tasks, described in the next section.

C. **Explanation Validation**: The explanations generated by the Explanation Assistant are evaluated by the application stakeholders to determine their suitability for the intended audience and their effectiveness with respect to the explanation goals specified in the Explanations Requirements.

This paper focuses on the technical phase (B) and describes it in full (Sec. 5). The surrounding socio-technical context involved in the requirement analysis (A) and explanation validation (C) is beyond the scope of the paper, as it may involve completely different contexts, e.g. legal, business, ethics, etc. In this paper, we assume that the explanation requirement analysis (A) has been done (see [17] for the details) and the resulting Explanations Requirements are available before the ETD phase (B) starts. They must enumerate all the explanations to be supported and, for each of them,

TABLE 1
Explanation Requirement `cs-1b` from the Credit Card Assessment application (Sec. 3.1 addressing a specific legal requirement. Some other (secondary) characteristics (e.g. explainability goals, trigger, priority) are omitted to save space. See [17] for a full example.

| Explanation ID | `cs-1b` |
|---|---|
| Legal Req. ID | DP.GDPR.8.3 (automated decision-making) |
| … | … |
| Minimum required content | Explanation of automated decision, the main data points used in the decision-making process |
| Audience | The data subject (i.e., the applicant) |
| Exemplar narrative | *"We made this decision with an automated scoring system that takes into account the information provided in your application as well as a credit score produced by a credit referencing agency."* |

must describe the following characteristics at the minimum:
- **Minimum required content**: the information required to be present in the explanation in order for it to satisfy the explanation's mandate.
- **Exemplar narrative**: text sentences crafted by the domain expert for a specific explanation.

Table 1 shows an example explanation requirement. The full list of characteristics is defined in [17]; the above is the minimum required to implement explanation requirements in the ETD phase.

## 5 EXPLANATION TECHNICAL DESIGN PHASE

This section describes the four tasks that constitute the ETD phase, which are overviewed in Fig. 2. From the provided Explanation Requirements and the application's Data Flows, the Data Engineer (1) models the audit trails of decisions made by the application, (2) creates queries to extract data from those audit trails as required by a particular explanation, (3) builds explanation plans to render the queried data in narratives suitable for end-user consumption, and (4) deploys the Explanation Assistant service for the specified application. Tasks 1–3 produce the Provenance Patterns (to record audit trails), the Provenance Queries, and the Explanation Plans to support all the explanations specified by the Explanation Requirements. Task 4 uses them to instantiate an Explanation Assistant service that produces explanations from the application's data. The steps in the above tasks are summarised with their input/output artefacts in Tables 3—6.

### 5.1 Task 1: Modelling the Provenance of a Decision

This task provides the means for the application to record the audit trails of its decisions, enabling us to trace back a decision to its input data and to identify the responsibility for each of the activities that led to the decision. Such an audit trail is also known as the *provenance of the decision*. Paraphrasing the definition of provenance by the World Wide Web Consortium (WC3) [34], we define the provenance of a decision as "a record that describes the people, institutions, entities, and activities involved in producing, influencing, or delivering" that decision. Such records provide the *history* of the decision and are a valuable source of data for generating explanations about what happened that

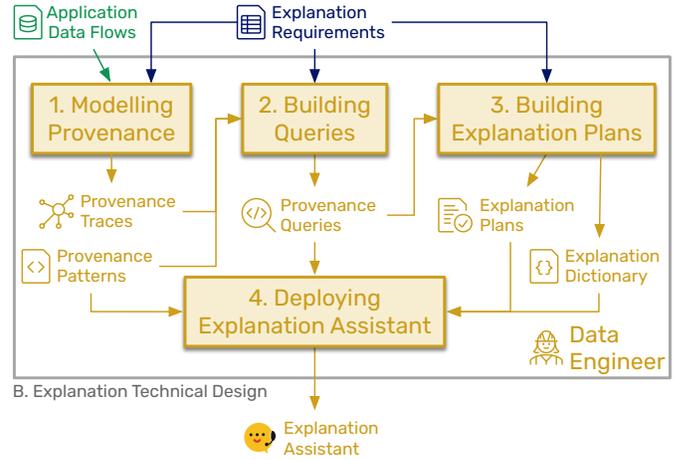

Fig. 2. The four tasks (numbered) in the Explanation Technical Design phase shown with the artefacts they produce or take as inputs.

led to the decision. In order to do so, however, the decision's provenance must contain sufficient details to support all the specified explanations, i.e. capturing all their minimum required content. In addition, it must satisfy the following *provenance requirements* to allow for retrieving the information needed to generate explanations in the next task [31]. The provenance must allow for:

**PR1** describing the various types of data of the universe of discourse, e.g. an application, an applicant, data, data providers, a decision, and so on;

**PR2** tracing back outcomes to their influencers;

**PR3** attributing or assigning responsibility to software systems or humans for actions or outcomes; and

**PR4** describing the various activities, their respective timing, and their contribution to outcomes.

Any provenance data model (e.g. PROV Data Model [34], Open Provenance Model [45]) or, indeed, knowledge graph (such as RDF graphs [46]) that supports the above requirements can be employed to model the provenance of a decision in the target application. With a chosen data model, the provenance of a decision can then be modelled following a modelling methodology such as PrIMe [33]. However, to support explanation generation via queries (Task 2), in addition to the above provenance requirements, the provenance to be recorded must follow some *predicted* patterns to allow for specific information to be located. Hence, this task must document the expected Provenance Patterns and share them with Task 2 to facilitate the construction of provenance queries. If provenance is recorded using the PROV Data Model, such patterns can be described using provenance templates [35]; or, if it is recorded as RDF graphs, the Shapes Constraint Language (SHACL) [47] could be employed.

Specifically, in this task (see Table 3), the Data Engineer inspects the Data Flows provided by the Application Stakeholder to locate the data entities therein that hold the minimum content required for *all* explanations (specified by the Explanation Requirements) (**PR1**, Step 1.1). From those data entities and with reference to the Data Flows, in Step 1.2, the Data Engineer determines how they contribute to a decision by the application; in many cases, this also involves identifying intermediary data entities and activities in the

6                                                                                                                                                                      EXPLAINABILITY-BY-DESIGN

TABLE 2
Glossary of the artefacts generated or used in the Explanation Technical Design phase.

| Artefact | Description |
| --- | --- |
| Application Scenario | A document describing the application, how it works, its context, its users, and decisions that it makes. |
| Data Flows | Diagrams or descriptions of the conceptual flows of data within the specified application, including some sample decisions and their associated application data. |
| Explanation Requirements | The socio-technical specification of explanations for the application, which enumerates all the explanations designed in EbD's Phase A and their characteristics. An example explanation requirement is shown in Table 1. |
| Explanation Assistant | A software service that provides explanation generation capability for the application for which it is configured. |
| Provenance Patterns | The set of patterns describing the shape of the provenance to be recorded for a specified application. |
| Provenance Traces | Full provenance traces of a number of sample decisions following the specified Provenance Patterns. |
| Provenance Queries | The set of queries to extract the relevant data from a decision's provenance to support explanation generation. |
| Explanation Plans | The syntax trees specifying the explanation narratives to be produced by the natural language generation engine. |
| Explanation Dictionary | A collection of application-specific linguistic configurations, including profiles of the supported target audience. |
| Test Explanations | Explanations produced by a configured Explanation Assistant from the application data provided with the Data Flows. |

TABLE 3
Task 1: Modelling Provenance

**What**: Constructing the provenance patterns to specify the audit trail to be recorded so that it contains all the data entities, activities, people, organisations, and systems that influence and lead to a decision, which are necessary to explain the decision.

**Inputs**: Explanation Requirements, Data Flows.

**Outputs**: Provenance Patterns, Provenance Traces.

1.1 Inspect the application's Data Flows and locate the data entities holding the minimum required content for *all* the specified Explanation Requirements.
1.2 Identify how the data entities found in Step 1.1 lead to the decision, the possible intermediary data entities and/or activities that connect them, and who or what is responsible for all those (i.e., the agents).
1.3 Capture all the above provenance elements and how one influences another in a set of Provenance Patterns.
1.4 Instantiate the Provenance Patterns with sample application data to ensure that the resulted sample Provenance Traces satisfy the provenance requirements and contain all the minimum content required by the Explanation Requirements.

TABLE 4
Task 2: Building Queries

**What**: Constructing the provenance queries for *each* explanation to extract the data to be included in the explanation's narrative.

**Inputs**: Explanation Requirements, Provenance Traces.

**Outputs**: Provenance Queries.

For each explanation requirement:
2.1 Locate the data elements required by an explanation's exemplar narratives in the sample Provenance Traces.
2.2 Construct a graph query that links a decision to all the identified data elements.

TABLE 5
Task 3: Building Explanation Plans

**What**: Constructing the NLG syntax tree for *each* explanation, defining the explanation's narrative to be generated.

**Inputs**: Explanation Requirements, Provenance Queries.

**Outputs**: Explanation Plans, Explanation Dictionary.

For each explanation requirement:
3.1 Construct the NLG syntax tree from the exemplar narrative of the explanation.
3.2 Replace the narrative's data elements with the corresponding variables from the provenance query for the explanation.
3.3 Specify the linguistic elements specific to each profile to be supported in the Explanation Dictionary.

TABLE 6
Task 4: Deploying the Explanation Assistant

**What**: Instantiating an Explanation Assistant service configured specifically to support the application's Explanation Requirements.

**Inputs**: Provenance Patterns, Provenance Queries, Explanation Plans, Explanation Dictionary, Data Flows.

**Outputs**: Explanation Assistant, Test Explanations.

4.1 Configure the Explanation Assistant service with the Provenance Patterns, Provenance Queries, Explanation Plans, and Explanation Dictionary produced in the previous tasks.
4.2 Test the instantiated Explanation Assistant with sample data from the Data Flows to produce Test Explanations.

flows and how they are linked to the decisions (**PR2**). Next, the activities that either use or generate the above data entities are identified (**PR4**). At the same time, the Data Engineer identifies who or what is responsible for those data entities and activities (**PR3**). The knowledge is then captured in a set of Provenance Patterns, typically one for each identified activity, to model the interactions between the influencers and the outcomes of that activity (Step 1.3). Finally, using the data of sample decisions provided with the Data Flows, the Data Engineer instantiates the Provenance Patterns to produce a set of Provenance Traces for those decisions (Step 1.4). This step aims to ensure that the resulting Provenance Traces satisfy all the provenance requirements (including that the decision is suitably linked to the influencers as per **PR1–4**) and contain the minimum information content required for all the specified explanations. See the PRIME methodology [33] for further guidance with this step.

## 5.2 Task 2: Building Queries

A *personalised* explanation is specific to a particular decision made by an application, and it must include the data, activities, and responsibilities specific to that decision-making. Thanks to the Provenance Modelling task, all such information should have been included in the provenance of a decision. However, the provenance itself is *not* an explanation for the decision and is also typically unsuitable to be presented to end users in its original form (e.g. an RDF graph or a PROV serialisation). The information or data required to explain a decision needs to be extracted from its provenance to construct an explanation that is personalised and targeted to the intended recipient (and in a form that is easy to digest). As the provenance (of a decision) can be represented as a graph, the required information contained within such a graph can be found therein by tracing back from the decision in question (**PR2**).

In Task 2 (Table 4), for *each* explanation specified in the Explanation Requirements, the Data Engineer examines the explanation's exemplar narrative and highlights the data elements therein that are specific to a particular decision. Those data elements are then identified in the sample Provenance Traces produced as a result of Task 1 from the Provenance Patterns (Step 1.4). Based on where the required data elements are found in the Provenance Traces, the Data Engineer constructs a graph query to describe how those elements are linked to the decision and, hence, can later be found in the decision's provenance trace (Step 2.2). A graph query language (e.g. SPARQL [48], Cypher [49]) is needed for this purpose. The query language must allow a query to select over provenance constructs, joining them via provenance relations, and filtering them by their attributes (e.g., data types, time, values) to identify the data required for constructing the explanation. An example is provided in Sec. 7.2.

## 5.3 Task 3: Building Explanation Plans

To construct an explanation that is easy to digest for end users, in this task, the Data Engineer builds Explanation Plans to dictate how a narrative is generated from the data elements retrieved for an explanation by its Provenance Queries. An explanation plan is a linguistic syntax tree, drawing its foundation from Natural Language Generation (NLG) pipelines [41], created for each sentence provided in an explanation's exemplar narrative (as specified by the Explanation Requirements). Such a sentence typically mentions data or information specific to a decision within otherwise static texts. For example, in the explanation about a credit card decision, the text "...your *credit score* produced by *credit referencing agency*..." contains 'credit score' and 'credit referencing agency', which are the data elements that can be queried from a decision's provenance, while the surrounding texts are unchanged with respect to application data from one decision to another. Hence, the syntax tree in an explanation plan does not only contain linguistic constituents but also references to variables to be bound to the results of the corresponding provenance query.

Explanations containing the same information may be presented to different audiences. For instance, the explanation that is included in the web page viewed by a credit card applicant may also be provided to a bank's support team when responding to the same applicant's enquiry over the phone. To support possible different profiles of usage, the *profiles* can be defined in an Explanation Dictionary to be shared by all the explanation plans constructed for a specific application. For example, from the same explanation plan, a 'customer support' profile may generate the phrase "...*the borrower's* credit score produced by credit referencing agency..." instead of "...*your* credit score" (as addressed to the applicant). When producing explanations, the application selects an *active* profile, which suitably configures the language generation of an explanation plan.

In summary, in Task 3 (Table 5), for *each* sentence in an explanation's exemplar narrative, the Data Engineer constructs a linguistic syntax tree (Step 3.1). The data elements specific to a particular decision in the syntax tree are then replaced by references to variables from the corresponding Provenance Queries (Step 3.2). For each profile of usage to be supported, relevant linguistic elements are to be defined in the application's Explanation Dictionary and referred to in the syntax tree (Step 3.3). The outputs of this task are, hence, a set of Explanation Plans for all the required explanations and the associated Explanation Dictionary. During the execution of an explanation plan, values from the corresponding provenance query (run over the provenance of a decision) are to replace variable references in its syntax tree. The 'instantiated' syntax tree will then be processed by a natural language generation (NLG) engine [41] to generate the final sentence to be presented to the target audience.

## 5.4 Task 4: Deploying the Explanation Assistant

The EbD methodology advocates for a reusable software component called *Explanation Assistant* service that operates in parallel with the main application and provides it with explanation capability via a set of well-defined Application Programming Interfaces (APIs). While Sec. 6 outlines the architecture of the service and the rationale behind the design, it is worth mentioning here that it should provide an API for the application to post application data required by the Provenance Patterns and logged from its execution to construct the provenance of a decision it makes. The application can then retrieve explanations for the specific decision via another provided API (see Fig. 3).

In this final technical task (Table 6), the Provenance Patterns, Provenance Queries, Explanation Plans, and Explanation Dictionary produced in the three previous tasks are bundled together to configure the Explanation Assistant service (Step 4.1). To ensure that the produced explanations satisfy the Explanation Requirements, in Step 4.2, via the Logging and Explanation APIs (Fig. 3), the Data Engineer employs the configured Explanation Assistant service to generate Test Explanations from the sample application data provided with the Data Flows. The Test Explanations are passed on to the next phase of EbD to be validated by the application stakeholders.

## 6 ARCHITECTURE

As previously alluded, we advocate that explanation capabilities are provisioned by a reusable Explanation Assistant service running in parallel with the main application making decisions requiring explanations. This service-based approach to provisioning explanations is beneficial in several ways:

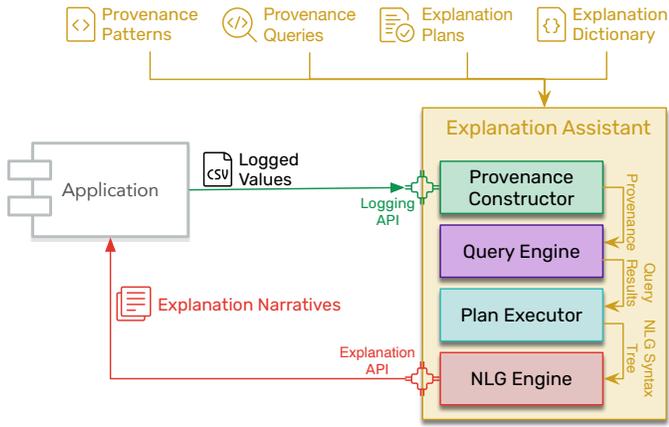

Fig. 3. Software architecture for an Explanation Assistant service.

- Separation of concerns: The generation of explanations is decoupled from the execution of the main application, allowing both to be independently maintained and run. Providing that the Data Flows of the application are fully specified and kept up-to-date, the application and the Explanation Assistant service can be developed and maintained by separate teams of engineers with their specific expertise.
- Maintenance of explanations: Requirements for explanation may evolve due to new regulations, changing business needs, or a mandate to support a new target audience (e.g., regulators). If the decision-making process of the application is unchanged, thanks to the separation of the explanation capability, the existing Explanation Plans, Explanation Dictionary, Provenance Queries, and Provenance Patterns can be amended, or new ones added, to support the new requirements without the need to update the main application.
- Reusability: The software components to support explainability (i.e., constructing provenance, querying provenance, natural language generation) can be shared across different applications; only their running configurations need to be application-specific. Potentially, the structure of those configurations can also be shared in some cases, expediting the explainability support for a new application (as later shown in Sec. 8.4).

In the above software architecture, shown in Fig. 3, the main application interacts with the Explanation Service via two APIs: (1) to log application data specified by the Provenance Patterns and (2) to retrieve specific explanations for a given decision. Whenever the application makes a decision, the data pertaining to the provenance of the decisions are posted to the Logging API (1) and stored by the Explanation Assistant. The provenance trace of the decision is only generated from the data according to the Provenance Patterns if and when needed. When certain explanations for a specific decision are required, the application requests the Explanation Assistant to produce them, via the Explanation API[5] (2), from the decision's previously logged data.

---

5. In Sec. 2 of the Supplementary Materials, we provide an example of such API calls (to the reference implementation of the Explanation Assistant) to illustrate how the API works in practice.

Within the Explanation Assistant service (Fig. 3), four main components work together in a pipeline to transform logged application data into the required explanation narratives. Their operations are to be configured by the Provenance Patterns, Provenance Queries, Explanation Plans, and Explanation Dictionary produced by ETD tasks 1–3. In particular, the Provenance Constructor is responsible for producing the full provenance of a decision from the data logged by the application according to the Provenance Patterns. Then, for each Explanation Plan to be executed on a decision, the Plan Executor runs the corresponding Provenance Query over the decision's provenance using the Query Engine. The executor takes the query results and instantiates the plan's syntax tree, replacing its variables with values retrieved from the provenance of the decision by the query. The executor also configures the syntax tree using the terms and linguistic features defined for the currently selected profile in the Explanation Dictionary. The completed NLG syntax tree is then passed on to the NLG Engine to produce the explanation narrative's sentence(s), which are returned to the main application. Note that we describe the roles of the components in a technology-agnostic manner since the technologies implementing their functionality depend on the chosen provenance data model. The following section provides more details on the specific technologies employed in our reference implementation of an Explanation Assistant service.

## 7 REFERENCE IMPLEMENTATION

We followed the tasks in the ETD phase to implement the explanations for the Credit Card Assessment (CCA) and School Allocation (SA) scenarios (Sec. 3). The implementation is available as an online demonstrator at https://explain.openprovenance.org with the introductory guidance provided in Sec 1 of the Supplementary Materials. Since personal data is used for the CCA and SA applications, real-world data could not be shared with us. Therefore, we opted to simulate the decision-making processes of the two applications based on the technical inputs and anonymised data provided by our collaborators, a credit referencing agency and a local education authority. The demoed applications generate fictitious data to feed into the simulated CCA/SA decision pipelines, resulting in (simulated) decisions to be explained. This section discusses the technology chosen for the reference implementation with respect to the Provenance Constructor, Query Engine, and Plan Executor (Fig. 3). In most cases, existing open-source technology components and industry standards were preferred over proprietary or non-standard options to expedite the implementation and facilitate future extensions. Due to the limited space, the technical description of the implementation cannot be fully included; however, examples are provided to illustrate how each component works. The full set of artefacts produced to support the explanations in both scenarios is available online at https://github.com/plead-project/EbD-artefacts.

### 7.1 Constructing Provenance of Decisions

In this implementation, we adopted the PROV data model (PROV-DM) [34], the *de jure* provenance standard by W3C, for modelling the provenance of a decision. The PROV data model defines three core types: *entity*, *activity*, and *agent*; which can be related to one another by PROV relations (**PR1**). In brief, provenance describes the generation and use of entities by some activities (**PR2**, **PR4**), which may be influenced in some ways by agents (**PR2**, **PR3**). In other words, PROV-DM provides a provenance vocabulary that satisfies all the provenance requirements to support explanation generation (Sec. 5.1).

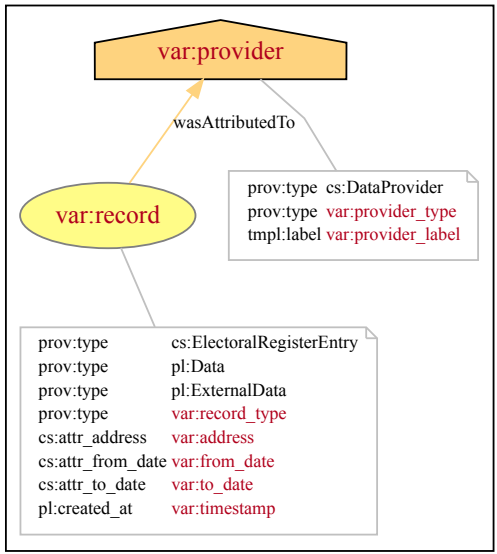

Fig. 4. A provenance template [35] describing that an entity identified by `var:record` was attributed to an agent identified by `var:provider`; the entity is of type `cs:ElectoralRegisterEntry` and the agent is of type `cs:DataProvider`.

Since it is not always possible or easy to instrument an (existing) application to directly record provenance records of sufficient quality, to facilitate the systematic capturing of the provenance of decisions, we adopt the PROV-Template approach [35] to *decouple* the modelling of the provenance and its construction. It allows the data engineer to describe the provenance information to be recorded in a set of *provenance templates* that contain provenance records with placeholders, called *variables*, to be filled in later by values logged by an application during its runtime, called *bindings*. In other words, those provenance templates implement Provenance Patterns (Task 1) using the PROV vocabulary. Fig. 4, for example, presents a simple provenance template describing that an electoral record, an entity, was attributed to a data provider, an agent. The identities of the actual record (`var:record`) and the provider (`var:provider`), i.e. the bindings for those variables, are to be logged by the application at run time. When the full provenance (of a decision) is required, all the provenance templates designed for the application are *expanded* by the Provenance Constructor, or instantiated, with the values of the corresponding bindings recorded by the application. The results are a set of expanded provenance records constituting the complete provenance of the decision. The Provenance Constructor employs the open-source ProvToolbox[6] library which provides provenance template expansion out-of-the-box.

6. ProvToolbox: http://lucmoreau.github.io/ProvToolbox/.

```
1  prefix pl <http://openprovenance.org/ns/plead#>
2  prefix cs <http://openprovenance.org/ns/
       ↪ creditscoring#>

4  select * from decision a prov:Entity

6  from derivation a provext:WasDerivedFromStar
7    join decision.id = derivation.generatedEntity

9  from record a prov:Entity
10   join derivation.usedEntity = record.id

12 where decision[prov:type] >= 'pl:Decision'
13 and record[prov:type] >= 'cs:CreditRecord'
14 group by decision aggregate record with Seq
```
Listing 1. A provenance query with the variables highlighted in red.

### 7.2 Querying Provenance

As the result of Task 1, the provenance templates can be expanded with application data logged at runtime to produce Provenance Traces, which are PROV provenance records. There are two key characteristics expected from a query in this context. First, the query should be able to return relevant PROV expressions. Second, it should be able to return a subgraph, typically linking a decision to one or more factors affecting it (**PR2**). From this subgraph, an explanation will be constructed (see Sec. 7.3) with the relevant extracted data points, such as names, dates, and so on, necessary to produce an explanation that is personalised to its recipient.

As existing graph query languages such as SPARQL [48] or Cypher [49] were designed to be domain-agnostic, they do not readily support the PROV data model. PROV relations, for instance, are not merely labelled edges in a graph but also contain associated data, such as the time an activity used a data entity. Therefore, a suitable query language must also be able to extract the "edges," or PROV-relation records, so that the data associated with them can be included in an explanation. SPARQL, while supporting path queries, cannot extract subgraphs; it only returns the start and end of such a path. Although Cypher and a proprietary extension of SPARQL[7] allow for subgraphs to be extracted, they can only work on a suitable graph representation and there is not yet a canonical graph conversion/encoding for PROV. Against this background, we extended the relational query engine LegoBase [50] to work directly with PROV records, used as the Query Engine, with which we defined a SQL-like query language to allow all PROV records to be selected and returned. The query language's definition is not in scope, but an example query is provided in Listing 1. We conjecture that such PROV queries can be compiled to Cypher.

In essence, a query language should allow for binding variables to both PROV types (e.g., Line 4) and relations (e.g., Line 6) using terms from PROV-DM; linking them via PROV relations through 'joins' (Lines 7, 10); and filtering the matched records by their attributes (Lines 12–13) to identify the right records required. Starting from a 'decision' (Line 4) of type `pl:Decision` (Line 12), the example query looks for all `prov:Entity` records (Line 9) of type `cs:CreditRecord` linked to it

7. See https://www.stardog.com/blog/a-path-of-our-own/.

via `prov:WasDerivationFrom`[8] record(s) (Line 6); all the records found are to be aggregated into a list (Line 14). Executed against the provenance trace of a decision, if the query matches, the query's variables (i.e. **decision**, **derivation**, **record**) are bound to the corresponding provenance records in the trace. Those matched provenance records and their properties will then be available for explanation plans in constructing sentences.

### 7.3 Executing Explanation Plans

To generate an explanation sentence, we employed the open-source SimpleNLG library [42] as the NLG Engine, which takes an NLG syntax tree and produces a grammatically correct sentence. Before that, the NLG syntax tree needs to be transformed from an Explanation Plan by the Plan Executor, which (1) processes references to variables returned by the corresponding Provenance Query and (2) applies the selected linguistic profile. Listing 2 shows an example Explanation Plan for the sentence "Your credit score was impacted by [records details]" (constructed for the CCA scenario). The sentence was broken up into the various part-of-speech components: the verb 'impact' (Line 2), the noun phrase 'credit score' (Lines 4–5) with a possessive specifier reference `##borrower-possessive` (Line 6), and the indirect object (Lines 7–31). The syntax tree also specifies other linguistic features of the sentence, such as this is a passive sentence in the past tense (Lines 32–33).

**Processing variables**. The '[records details]' part of the above sentence refers to data that is specific to a particular decision and, hence, requires the result of the associated provenance query (Listing 1) after being executed over the decision's provenance. To support the (potentially) multiple instances of PROV records returned by the query and to represent them in a user-friendly format, several auxiliary functions can be employed directly in an Explanation Plan. For instance, an iterator (Line 12) is used to connect the representation of all the PROV records bound to the variable **record** (Line 14) in the sentence. For each of the records, the value of its `prov:type` property is retrieved and looked up in an Explanation Dictionary (Lines 17–22) to be included in the sentence as a noun phrase, along with its identifier (Lines 25–29). This allows for sentences such as "Your credit score was impacted by your salary (records/960) and a late payment (records/956)". In addition (not shown in the listing), suitable HTML markups can be specified in the Explanation Plan to present record identifiers as hyperlinks, making an explanation navigable and allowing its recipients to drill down into its contents if interested.

**Applying a profile**. The explanation sentence can be further customised to suit different 'profiles' of the target audience. For instance, explanations for a credit card decision could be presented to two different audiences: the applicant and the bank's staff. These two profiles are defined in an Explanation Dictionary with relevant application-specific terms for each of them. In the above example, the reference `##borrower-possessive` (Line 6) marks an entry to be looked up from the application's Explanation Dictionary. It will

---

8. We also extended PROV terms in the query language to support querying the transitive closure of `prov:WasDerivationFrom` relations by querying for the `provext:WasDerivedFromStar` relation.

```json
{ "type": "clause",
  "verb": "impact",
  "object": {
      "type": "noun_phrase",
      "head": "credit_score",
      "specifier": "##borrower-possessive" },
  "indirect_object": {
      "type": "clause",
      "object": {
          "type": "coordinated_phrase",
          "conjunction": "and",
          "@iterator": {
              "type": "@iterator",
              "@variable": "record",
              "@clause": "coordinates",
              "@element": {
                  "type": "@funcall",
                  "@object": "record",
                  "@property": "prov:type",
                  "@function": "lookup-type",
                  "@arg1": "noun_phrase",
                  "@arg2": "csd",
                  "post-modifiers": [{
                      "type": "adjective_phrase",
                      "head": {
                          "type": "@funcall",
                          "@object": "record",
                          "@field": "id",
                          "@function": "noun+localname" }
                  }] } } },
      "complementiser": "by" },
  "features": { "type": "features",
      "tense": "past", "passive": "true" }
}
```

Listing 2. An Explanation Plan (represented in JSON) to produce a sentence like "Your credit score was impacted by missed payment (*xyz/nn*)."

be converted into a second-person possessive determiner ("your") if the former profile is selected or a third-person possessive determiner ("their") if the latter is selected.

## 8 ASSESSING THE ETD PHASE

Assessing the costs and benefits of a software engineering methodology is not always straightforward and often imprecise. It is usually through multiple applications of a methodology in various contexts that its key characteristics can be adequately understood. Moreover, there is currently no alternative software engineering approach to addressing explanation requirements against which we could compare ours. Nevertheless, with the CCA and SA scenarios (Sec. 3), we can already execute the ETD phase to produce explanations for them and preliminarily assess its qualities in a case study [44, Ch. 5]. In this section, we present the case study design and discuss the evidence observed from the experience of the above execution to inform potential adopters of the methodology.

### 8.1 Case Study Design

The **objective** of our assessment is to evaluate and characterise the technical efforts required in the ETD phase to support explanations *with respect to* some Explanation Requirements (specified in the Explanation Requirement Analysis phase). It can be addressed by answering the following **research questions**.

**RQ1** How much effort does the ETD phase require to support explanations?

**RQ2** Are there reusable elements that help reduce the technical costs of supporting explanations?

The case to be studied is the execution of the ETD phase over two application scenarios. Hence, it is an *embedded* case study with two units of analysis (UA) [44]: CCA and SA scenarios. In more detail, a legal expert had previously carried out the Explanation Requirement Analysis for the scenarios, providing us with a set of Explanation Requirements for each of them (see Table 1 for an example). Each requirement specifies a narrative to be produced from a decision's provenance, consisting of one or more sentences. From the requirements, we executed the ETD tasks to produce the specified explanations for decisions in the CCA and SA scenarios. Table 8 provides the numbers of Explanation Requirements and sentences to produce in each scenario.

**Collection of data**. To explore RQ1, we studied the artefacts produced by the ETD phase in each UA and the total development time incurred. For RQ2, we identified artefacts and technical components that were reused during the ETD tasks. The data was collected retrospectively from the Git repository that keeps track of the edit history of all the produced artefacts (e.g., provenance templates, queries, explanation plans).

**Limitations and Scope**. First, the case study is limited to the execution of the ETD phase only, i.e. producing explanations from the input Explanations Requirements and the Data Flows, and we simulated the decision-making pipelines; the effort to integrate the resulting Explanation Assistant service with the respective decision-making applications is not in scope. Second, adopting the software architecture described in Sec. 6, we implemented a *reusable* Explanation Assistant service (as detailed in Sec. 7). The service's implementation effort is not in scope since it is not part of the ETD phase. Third, three of the authors were the data engineers who executed the ETD tasks in the two UA. Given their intimate knowledge of the tasks, they must be viewed as engineers who are experienced with such tasks. Having said that, because the data were collected retrospectively from a Git repository after all the development had concluded, the method to collect data was not intrusive; the engineers were just making use of their day-to-day tools with no interruption from the study. Fourth, the reported development time was not directly measured but estimated from the timestamps of each UA's first and last commits. The execution of the UA did not overlap in time. Finally, our cost evaluation assumes that the generated explanations are suitable and have been validated separately by stakeholders.[9]

In the following subsections, we answer the research questions by analysing the collected data: for RQ1, Sec. 8.2 focuses on the number of produced artefacts and their complexity while Sec. 8.3 discusses the incurred development time; Sec. 8.4 analyses instances where artefacts were reused and the potential impacts with respect to RQ2.

---

[9]. While the validation of explanations is not in the paper's scope, we can report that we ran focus groups and individual interviews to receive feedback on the produced explanations (8 people from the credit referencing agency and 4 from a school admissions team at a local education authority). The interviewees were overwhelmingly positive about the usefulness of the produced explanations.

TABLE 7
Cost metrics of ETD tasks

| Task | Cost |
| --- | --- |
| Provenance Modelling | • Total number of provenance records |
| Building Queries | • Total number of joins & filters in queries |
| Building Explanation Plans | • Total number of nodes in all syntax trees<br>• Numbers of profiles and dictionary terms |

## 8.2 RQ1: Technical Products and their Complexity

To help understand the effort involved in developing explanations for the two selected scenarios (RQ1), we propose a number of metrics that reflect the cost of the four ETD tasks based on the number of resulting technical products, i.e. artefacts, in each task and their complexity (summarised in Table 7):

- Provenance Modelling: The effort of this task can be measured by the number of provenance templates produced and their complexity, defined by the total number of provenance records contained therein.
- Building Queries: This task produces Provenance Queries; its cost can be measured by the number of queries and their complexity. The latter is approximated by the total number of 'joins' and filtering conditions employed in the queries.
- Building Explanation Plans: An Explanation Plan consists of a syntax tree whose complexity can be quantified by the number of linguistic elements they contain, i.e., the total number of nodes within its narrative's syntax trees. This task's effort is also affected by the number of profiles supported in the Explanation Dictionary, reflected in the number of dictionary terms defined therein.
- Deploying Explanation Assistant: The reference Explanation Assistant service (Sec. 7) is a prebuilt software which does not require further development. In this task, the Explanation Assistant service was configured using the artefacts produced by the above tasks and the effort was minimal.

To provide an indication of the effort involved, Table 8 shows the numbers of artefacts produced to support the explanations in the case study's two UA. They have 7 and 4 input Explanation Requirements, respectively. The explanation narrative for an explanation requirement may consist of multiple sentences; those sentences may be divided into multiple explanation plans to facilitate their construction. We, therefore, also report the total number of explanation sentences produced for each scenario. Further detailed cost measurements are provided in Sec. 3 of the Supplementary Materials.

## 8.3 RQ1: Development Time

A team of three engineers, E1, E2, and E3, carried out the ETD tasks for the two scenarios. Their roles were as follows:

**E1** spent 50% of their time on this project and was responsible for crafting the explanation plans (constructing the NLG syntax trees, i.e., Task 3).

**E2** spent 25% of their time on this project and was responsible for modelling the provenance for the two scenarios (Task 1) plus writing the provenance queries (Task 2).

TABLE 8
The number of artefacts produced to support the explanations requirements of the CCA and SA scenarios with their average complexity in brackets.

| Artefact | CCA | SA |
| --- | --- | --- |
| Explanation Requirements | 7 | 4 |
| Sentences | 34 | 12 |
| Provenance Templates | 11 [5.8] | 7 [7.7] |
| Provenance Queries | 11 [7.6] | 4 [16.3] |
| Explanation Plans | 22 [25.3] | 11 [24.1] |
| Profiles | 2 | 2 |
| Dictionary Terms | 16 | 10 |

TABLE 9
The estimated engineer time in person days incurred to support the explanations for the CCA* and SA† scenarios.

| Metric | CCA | SA |
| --- | --- | --- |
| Total development time in person days | 29.6* | 3.2† |
| Development time per sentence in hours | 6.5* | 2† |

* including Explanation Assistant co-development time
† *no co-development was needed*

**E3** spent 5% of their time on this project as an advisor, training and supporting E1 by providing a few initial example explanation plans to help them get started.

Since any changes to the artefacts produced for the two scenarios were tracked in a Git versioning system, we can plot the involvement of the three engineers onto two timelines, which are provided in Sec. 4 of the Supplementary Materials (due to the limited space here). Based on the timespan between the first and the last Git commits and the level of effort reported above, we estimated the total development time incurred for each scenario, summarised in Table 9.

It should be noted that the development of explanations for the CCA scenario started when the technical support for some provenance queries and explanation plans had not been fully implemented. Therefore, there were occasions when some tasks had to wait for the required functionality to be added. As a result, the time estimation for the CCA scenario includes part of the development time of the reference Explanation Assistant service. In contrast, the explanations for the SA scenario were developed after all the required technical support had been in place. Given the above context, we believe that the development time of the SA scenario (Table 9) was more indicative of the required effort to carry out the ETD tasks. Considering that there were 12 sentences to produce in the SA scenario and 7.5 working hours a day, we estimate that it took about 2 hours on average to support one explanation sentence. This should be understood as a real upper bound since no GUI-based tools were available, and the engineers relied on simple command line interfaces to carry out the work.

### 8.4 RQ2: Reusability

In both UAs, the numbers of Provenance Queries are significantly lower than the corresponding numbers of Explanation Plans: 22 plans were supported by 11 queries in the CCA scenario, and 11 by 5 in the SA scenario (see Table 8). The reason is that many of the queries were shared between multiple explanation plans. For instance, one sentence may expand on the information provided by another sentence by providing more details that are available in the result of the same query used by that sentence. Sec. 5 in the Supplementary Materials provide more details on how queries were reused.

In addition, the profiles of the target audience, in many cases, can be easily modified to be used in a new application. In fact, the two profiles crafted for the CCA scenario to address explanations to (1) the borrower and (2) the bank staff were copied to use for the SA scenario to address explanations to (1) the applicant and (2) the organisation staff. The main modifications required were changing the words 'borrower' to 'guardian' and 'the bank' to 'the school'.

Finally, with its well-defined APIs, the reference Explanation Assistant service was reusable between the two implemented scenarios. When developing the explanations for the SA scenario, we only needed to create the relevant artefacts (Table 8) from the input Explanation Requirements; no further development of the Explanation Assistant service was required.

The above evidence demonstrates there is a high potential for reusability of the products from the ETD phase. For instance, existing Provenance Queries and Explanation Plans could be extended to support new Explanation Requirements. Moreover, an organisation could even repurpose existing Explanation Plans to support similar, common Explanation Requirements (e.g., concerning data protection) for a new application.

**Iterative Development**. During the study, we additionally observed that sometimes the data needed in an explanation plan was not readily available from its corresponding provenance query. This then necessitated the revising of the query to retrieve the required data. However, in some cases, the missing data was not captured in the provenance traces; and to fix this, some provenance templates were revised to capture the extra data from the application. In other cases, a mistake identified in a query was fixed, resulting in edits in explanation plans depending on the query. Inspecting the development history of provenance templates, queries, and explanation plans (recorded in the git repository for the above artefacts), we identified 13 occasions of such iterative revisions, 7 of those involving edits by two engineers (see Sec. 6 in the Supplementary Materials). Therefore, given such an iterative nature of the ETD tasks, to carry out the ETD phase effectively, close collaboration within the development team will likely be essential.

## 9 CONCLUSIONS

The increasing dependence of decision-making on some level of automation has naturally led to discussions about the trustworthiness of such automation, resulting in calls for transparent automated decision-making and making such systems explainable. This article envisages Explainability-by-Design, a methodology to enable organisations to address their explainability systematically and holistically: (A) analysing explanation requirements within a social/legal context, (B) implementing the requirements technically, and (C) validating the produced explanations. We have described the Explanation Technical Design phase (B) and presented a software architecture for a reusable Explanation

Assistant service to provision explanation capability. An online reference implementation of such a service is available as a demonstrator. Following the methodology, organisations will be able to design their decision-making systems to track the provenance of decisions and produce individually tailored explanations for them to meet legal, social, or business requirements. In addition, we discussed various cost metrics associated with the methodology measured from the enactment of the methodology over two application scenarios in a case study. We have shown that thanks to the well-delineated steps and a reusable Explanation Assistant service, the approach requires modest development time per explanation sentence to be generated.

While the two chosen scenarios drew their explanation requirements from the regulatory requirements for explaining credit card and school allocation decisions, the methodology is not obligated to be legally driven. In fact, explanation requirements could equally arise from business requirements (e.g., to improve customer trust, to reduce customer enquiries, to support internal audits) or somewhere else. In machine planning, for instance, which is now used in safety-critical applications (such as shipping, oil-well drilling, and controlling swarms of unmanned vehicles), explaining why a certain action was chosen over another in a plan is key in engendering trust in the users who are responsible and accountable for authorising the execution of a machine-generated plan [51].

The presented EbD's technical workflow (the ETD phase) can be extended in a number of ways. Notably, it does not attempt to explain the inner working of algorithmic models employed by decision-making systems. Instead, it takes a holistic approach to produce explanations about the data and processes surrounding such models. At the same time, it does not preclude using existing XAI techniques to explain a recommendation by an ML model such as LIME [13] or SHAP [52]. Such techniques would complement and augment explanations about the decision-making pipelines supported by EbD by drawing out the relations between the input data and a model's recommendation in a particular decision. The provenance of a decision could then be extended to include those relations, allowing provenance queries to pinpoint the input data that was influential and, by so doing, making the explanations even more specific.

Although the ETD phase was designed to produce ex-post explanations about what happened, its technical approach can be extended to produce counter-factual explanations to provide insights into which external factors could be changed to arrive at a desired outcome [53]. In essence, a system can simulate counter-factual cases and produce the corresponding (hypothetical) decisions, with their provenance recorded as in an actual case. The provenance traces of the counter-factual cases can then be summarised or compared with that of the actual decision to generate counter-factual explanations. An initial exploration of the idea was included in the explanations of loan decisions reported in [31].

The reference implementation of explanations for the Credit Card Assessment and School Allocation scenarios was done without any tooling support available. The reported development time would have been significantly reduced if software tools were available to support the developers in their tasks. Such tools could help with visually crafting Provenance Queries (Task 2) and linguistic syntax trees (Task 3). They could also check whether the Provenance Traces are all connected (**PR2**, Task 1) and if the variable references in Explanation Plans match those available in Provenance Queries.

Finally, we have demonstrated that the Explanation Assistant service, as a reusable system, played a critical part in keeping the ETD phase's execution time low for the two studied scenarios. The methodology is, however, independent of the actual provenance templating system, the provenance query languages, and the explanation planning engine. To increase the reusability of the approach further, constructing well-defined interfaces between the service's components would allow alternative templating engines, query languages, and explanation planners to be plugged in to enable some specific functionality.

## ACKNOWLEDGMENTS

The work presented in this article has been supported by the UK Engineering and Physical Sciences Research Council (EPSRC) via the Grants [EP/S027238/1] and [EP/S027254/1] for the PLEAD project, [EP/R033722/1] for the THuMP project, and [EP/V00784X/1] for the Trustworthy Autonomous Systems Hub.